\documentclass[prl,twocolumn,showpacs,amsmath,amssymb,superscriptaddress,floatfix]{revtex4}

\usepackage[english]{babel}
\usepackage{graphicx}
\usepackage{times}

\begin{document}

\title{Fermi-liquid versus non-Fermi-liquid behavior in triple quantum dots}

\author{Rok \v{Z}itko}
\affiliation{J. Stefan Institute, Ljubljana, Slovenia}

\author{Janez \surname{Bon\v ca}}
\affiliation{Faculty of Mathematics and Physics, University of
Ljubljana, Ljubljana, Slovenia}
\affiliation{J. Stefan Institute, Ljubljana, Slovenia}

\date{\today}

\begin{abstract}
We study the effect of electron hopping in triple quantum dots
modelled by the three-impurity Anderson model. We determine the range
of hopping parameters where the system exhibits the two-channel Kondo
effect and has non-Fermi-liquid properties in a wide temperature
interval. As this interval is entered from above, the conductance
through the side dots increases to a half of the conductance quantum,
while the conductance through the system remains small. At lower
temperatures the conductance through the system increases to the
unitary limit as the system crosses over to the Fermi-liquid ground
state. Measuring the differential conductance in a three terminal
configuration provides an experimental probe into the NFL behavior.
\end{abstract}


\maketitle

\newcommand{\vc}[1]{\boldsymbol{#1}}
\newcommand{\ket}[1]{|#1\rangle}

Quantum impurity models describe interaction between a point-like
impurity with internal degrees of freedom and a continuum.  The
Kondo model of a magnetic impurity accounts for the screening of
the impurity spin with decreasing temperature (the Kondo effect).
Generalized Kondo models, such as the two-channel Kondo (2CK)
model and the two-impurity Kondo (2IK) model, display NFL physics
and quantum criticality, which are the central paradigms used to
interpret the unusual behavior found in some systems at low
temperatures.  The 2CK model may explain unusual logarithmic
temperature dependence of the magnetic susceptibility and linear
vanishing of the quasiparticle decay rate in some Ce and U
compounds at low temperatures \cite{cox1998}. The 2IK model was
applied to study the competition between magnetic ordering and
Kondo screening, which determines the ground state of the
heavy-fermion compounds \cite{jones1988, affleck1995}.

Quantum dots provide tunable mesoscopic realizations of quantum impurity
models \cite{wiel2002, craig2004}, where the relevant model is usually the
ordinary single-channel Kondo or Anderson model \cite{pustilnik2004}. The
magnetic moment is either fully screened for a spin $1/2$ impurity with
Fermi-liquid (FL) ground state or underscreened for a $S>1/2$ impurity with
a singular FL ground state \cite{koller2005, mehta2005}. Recently, several
experimental realizations of the 2CK model using quantum dots have been
proposed \cite{oreg2003, anders2004, pustilnik2ck2004, bolech2005}. In the
system of three quantum dots -- a small dot embedded between two larger dots
-- the channel asymmetry can be tuned to small values and NFL behavior was
predicted in a limited temperature range \cite{kuzmenko2003}.

We consider two related systems of three Anderson impurities coupled in
series between two conduction channels. We analyze new behavior that results
from the presence of two equivalent screening channels (as in the 2CK model)
combined with either two-stage Kondo screening \cite{wiel2002} or magnetic
ordering \cite{craig2004}, which both lead to a single uncompensated spin at
intermediate temperatures. If the impurities are coupled only by exchange
interaction, the system has a NFL ground state of the 2CK type with a
residual $\ln 2/2$ zero-temperature entropy \cite{affleck2ck1992, cox1998}.
In the experimentally relevant case where the exchange interaction is
generated by the superexchange mechanism due to electron hopping,
the channel symmetry is broken and the system is described by the
asymmetric 2CK model. We analyze the parameter ranges where NFL
behavior is exhibited and identify the regime where the experimental
observation is most likely.

{\it Models.} -- We study the three-impurity models described by the
Hamiltonian $H=H_\mathrm{b} + H_\mathrm{imp} + H_\mathrm{c}$,
where $H_\mathrm{b} = \sum_{\nu, k,\sigma} \epsilon_k c^\dag_{\nu k
\sigma} c_{\nu k \sigma}$ describes the left and right conduction lead
($\nu=L,R$) and
$H_\mathrm{c} = \sum_{k\sigma} V_k ( c^\dag_{Lk\sigma} d_{1\sigma} +
c^\dag_{Rk\sigma} d_{3\sigma} + \text{H.c.} )$ describes the coupling
of the bands to the left and right impurity (numbered 1 and 3, while 2
is the impurity in the middle).
$H_\mathrm{imp}$ is either the Hubbard Hamiltonian (model I)
\begin{equation}
H_\mathrm{imp}^\mathrm{I} = \sum_{i=1}^3 \frac{U}{2} (n_i-1)^2 \\
+ \sum_{i=1}^2 \sum_\sigma t \left( d_{i\sigma}^\dag d_{i+1,\sigma} +
\text{H.c.} \right),
\end{equation}
where $U$ is the on-site Coulomb electron-electron repulsion, $n_i =
\sum_\sigma d^\dag_{i\sigma} d_{i\sigma}$ is the electron number on
site $i$ and $t$ is the inter-impurity hopping, or the exchange-only
variant of the former (model II)
\begin{equation}
H_\mathrm{imp}^\mathrm{II} = \sum_{i=1}^3 \frac{U}{2} (n_i-1)^2 + J \vc{S}_1
\cdot \vc{S}_2 + J \vc{S}_2 \cdot \vc{S}_3,
\end{equation}
where $\vc{S}_i = 1/2 \sum_{\alpha\beta} d^\dag_{i\alpha}
\boldsymbol{\sigma}_{\alpha\beta} d_{i\beta}$ is the spin operator on
site $i$ [$\boldsymbol{\sigma}$ is the vector of Pauli matrices], and
$J$ is the exchange constant. We set $J$ to the superexchange value of
$J=4t^2/U$ to relate the two models for $t \ll U$. Both models are
particle-hole (p-h) symmetric. Replacing spin exchange interaction with
hopping enables charge transfer between the channels and induces
channel asymmetry \cite{kuzmenko2003, zarand2006} which drives the
system to a Fermi-liquid ground state \cite{cox1998}.

In model I three different regimes are expected as $t$ is decreased
\cite{tripike}: molecular-orbital (MO) Kondo regime, antiferromagnetic
spin-chain (AFM) Kondo regime and two-stage Kondo (TSK) regime.
In MO regime, two electrons occupy bonding molecular orbital, while the
third electron in non-bonding orbital develops local moment which is Kondo
screened.
In AFM regime, three on-site local moments bind at $T \sim J = 4t^2/U$ into
a rigid antiferromagnetic spin-chain with total spin $1/2$; this is followed
by the screening of the collective spin.
In TSK regime, the moments are quenched successively: on left and
right dot at the upper Kondo temperature $T_K^{(1)}$, while on middle
dot at an exponentially reduced lower Kondo temperature $T_K^{(2)}$
\cite{tripike, cornaglia2005tsk, vojta2002, sidecoupled}. All three
regimes become qualitatively similar at sufficiently low temperature:
the remaining degree of freedom is one spin $1/2$ coupled to two Fermi
liquids.
The most general effective Hamiltonian describing model I that is allowed by
the symmetries is the 2CK model with broken channel symmetry. Between TSK
and AFM regimes, we identify a wide crossover region where NFL behavior is
experimentally most accessible.

Model II has AFM and TSK regimes separated by the crossover
regime. There is clearly no MO regime; instead, the AFM regime extends
to the region of high $J$, where the two models describe very
different physical systems. Since the left and right conduction
channels are not communicating (in the sense that
there are no $L \leftrightarrow R$ cotunneling processes), the channel
symmetry is maintained and a stable 2CK NFL ground state is expected
for all $J$. In both models (I and II) NFL behavior sets in at the
highest temperature in the crossover regime.

\begin{figure}
\includegraphics[width=8cm,clip]{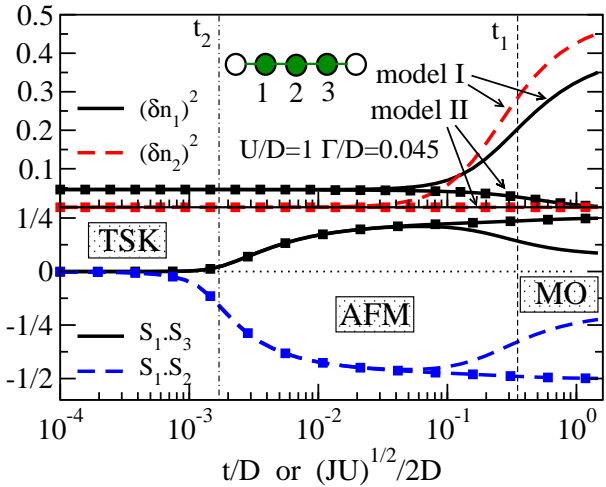}
\caption{(Color online) Charge fluctuations and spin-spin correlations of
model I (lines without symbols) and model II (lines with symbols) as a
function of the inter-dot coupling $t$ (for model I) or corresponding
$J=4t^2/U$ (for model II). The molecular-orbital (MO) regime is
characterized by large on-site charge fluctuations, the antiferromagnetic
spin-chain regime (AFM) by negative spin correlations of neighboring spins
1-2 and positive correlation of spins 1-3, and the two-stage Kondo regime
(TSK) by vanishing spin correlations.
}
\label{fig_c1}
\end{figure}

{\it Results.} -- We performed calculations using the numerical
renormalization group (NRG) method ($\Lambda=4$)
\cite{wilson1975}. We assumed a constant density of states
$\rho_0=1/(2D)$, where $2D$ is the band-width, and a constant
hybridization strength $\Gamma=\pi \rho_0 |V_{k_F}|^2$. In
Fig.~\ref{fig_c1} we show the ground state expectation values of
charge fluctuations $(\delta n_i)^2 = n_i^2 - \langle n_i
\rangle^2$ and spin-spin correlations between neighboring
$\vc{S}_1 \cdot \vc{S}_2$ and between side impurities $\vc{S}_1
\cdot \vc{S}_3$. For model I, the smooth cross-over from MO to AFM
regime, predicted to occur on the scale of $t_1 \sim U/2\sqrt{2}
\approx 0.35 D$ \cite{tripike}, is reflected in the decrease of
charge fluctuations and the increase of spin-spin correlations.
The cross-over from AFM to TSK regime occurs when $J \sim
T_K^{(1)}$ or $t_2 \sim 1.7\ 10^{-3} D$: as $t$ decreases past
$t_2$ the spin-spin correlations tend toward zero as the spins
decouple.  For model II, the results in the TSK regime match
closely those of model I, while in the AFM regime near $t\sim t_1$
the differences become notable.  Large values of $J\gg \Gamma$
suppress charge fluctuations on side-dots, $(\delta n_1)^2\to 0$,
while local moments on impurities tend to form a well developed
AFM spin-chain (for comparison, in isolated three-site spin chain
$\langle \vc{S}_1 \cdot \vc{S}_2 \rangle = -1/2$ and $\langle
\vc{S}_1 \cdot \vc{S}_3 \rangle = 1/4$).

\begin{figure}
\includegraphics[width=8cm,clip]{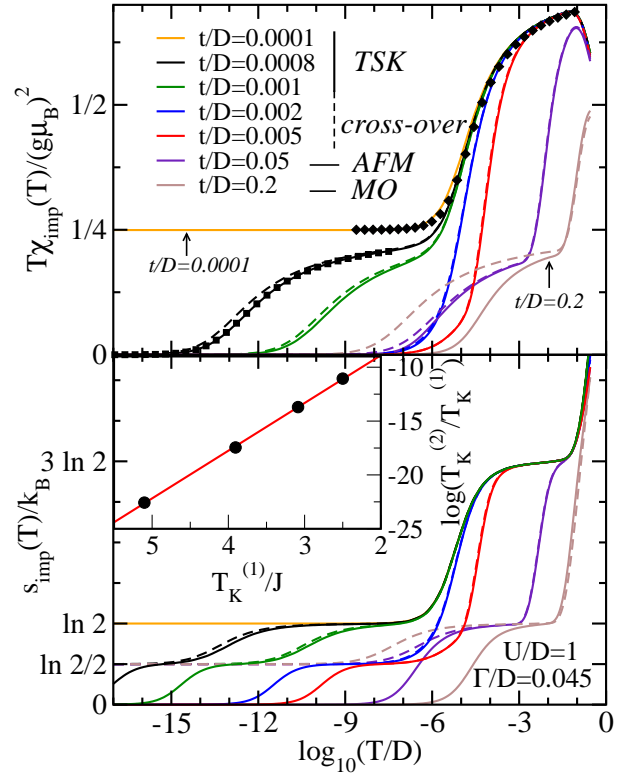}
\caption{(Color online) Impurity susceptibility and entropy for model I 
(full lines) and model II (dashed line) with $J=4t^2/U$.
Lozenges $\blacklozenge$ are a fit to Bethe-Ansatz results for 
one-channel Kondo model (multiplied by two and shifted by $1/4$) and
squares $\blacksquare$ are a fit to NRG results for 2CK model. {\it Inset:}
$T_K^{(2)} = a T_K^{(1)} \exp(-b T_K^{(1)}/J)$ scaling of the
second Kondo temperature of model I. $a=0.97$, $b=4.4$,
$T_K^{(1)}/D \approx 1.0\ 10^{-5}$. }
\label{fig_d1}
\end{figure}

\begin{figure}
\includegraphics[width=8cm,clip]{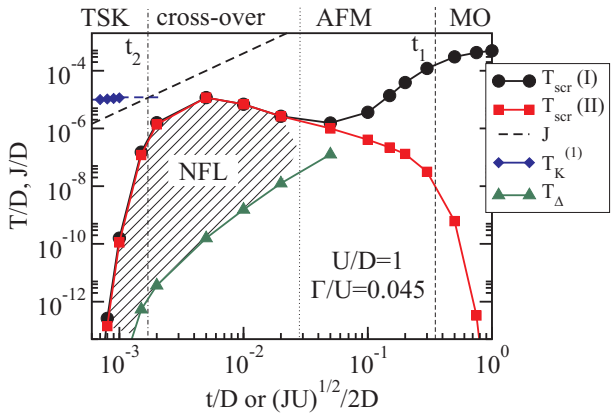}
\caption{(Color online)
Cross-over scales of models I and II as functions of
the inter-dot coupling. The magnetic screening temperature
$T_\mathrm{scr}$ is defined by $T_\mathrm{scr}
\chi(T_\mathrm{scr})/(g\mu_B)^2=0.07$; it is equal to the Kondo
temperature when screening is due to the single-channel Kondo
effect.  $T_\Delta$ is here defined as
$s_\mathrm{imp}(T_\Delta)/k_B=\ln2/4$.
}
\label{fig_tk}
\end{figure}

In Fig.~\ref{fig_d1} we plot the impurity contribution to
susceptibility and entropy. The ground state of model I is
nondegenerate, $s_\mathrm{imp}=0$, and the impurities are fully
screened for all $t$. In MO regime the system undergoes single-channel
Kondo screening with $T_K$ that increases with $t$ and becomes
constant for $t \gg U$, see Fig.~\ref{fig_tk}.
In AFM regime, the binding of spins is most clearly discernible in the
curves calculated at $t/D=0.05$ which show a kink in $s_\mathrm{imp}$
at $3\ln 2$ (local moment formation), followed by an exponential
decrease to $s_\mathrm{imp}=\ln2$ at $T \sim 4t^2/U$. The Kondo
screening in AFM regime is of single-channel type for $t/D \gtrsim
0.02$. Between $t/D=0.02$ and $t_2$ there is a {\it cross-over regime}
with NFL-like properties. Here magnetic ordering competes with the
single-channel Kondo screening of left and right impurity. The
magnetic moment is rapidly quenched at $T \sim T_\mathrm{scr} \sim J$,
yet the entropy does not go to zero but exhibits a $\ln2/2$ NFL
plateau. At still lower temperature $T_\Delta$, NFL fixed point is
destabilized by the channel asymmetry and the system crosses over to
the FL ground state characteristic of the conventional Kondo
model. Note that in this regime $T_\mathrm{scr}$ is high while
$T_\Delta$ is low (Fig.~\ref{fig_tk}), making this range suitable for
experimental study of NFL physics.

In TSK regime, the left and right impurity are screened by the
single-channel Kondo effect at temperature $T_K^{(1)}$ that is nearly
the same for all $t \lesssim t_2$ (Fig.~\ref{fig_tk}).  The central
impurity is screened by the 2CK effect at $T_K^{(2)}$, below which the
system is near the NFL fixed point with $\ln 2/2$ entropy. In the
inset to Fig.~\ref{fig_d1} we show that $T_K^{(2)}$ scales as
$T_K^{(2)} \propto T_K^{(1)} \exp( -b T_K^{(1)}/J )$, as expected for
the TSK effect \cite{vojta2002, cornaglia2005tsk, sidecoupled}.

Model II has a stable NFL ground state. For low $J$, it has a TSK
regime where the the Kondo temperature $T_K^{(2)}$, determined by $J$,
is lower than that of the corresponding model I, set by
$\max\{J_1,J_2\}=J_1>J$ (Fig.~\ref{fig_tk}).
In the {\it crossover regime} physical properties of model II for
$T>T_\Delta$ match closely those of model I. In AFM regime, the Kondo
temperature is a non-monotonous function of $J$. The energy required
to break the doublet spin-chain state increases with $J$ and the
effective Kondo exchange constant $J_K \propto \Gamma/J$ decreases.
$T_K$ therefore decreases exponentially with increasing $J$.

\begin{figure}
\includegraphics[width=8cm,clip]{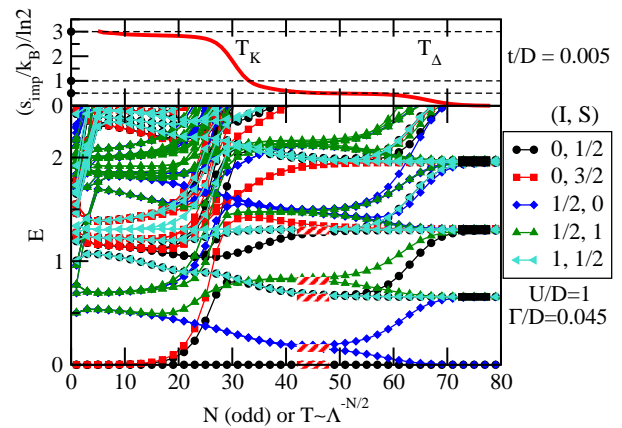}
\caption{(Color online) NRG eigenvalue flow
of model I ($\Lambda=2$, $z=1$) in the cross-over regime (below) and the
corresponding impurity entropy (above).
The states are classified according to the total isospin and total spin
quantum numbers, $(I, S)$.  Black full strips correspond to FL spectrum.
Gray (red online) hatched strips are at energies (after rescaling
by 1.296) $1/8$, $1/2$, $5/8$, $1$ \cite{affleck2ck1992}.
}
\label{fig_i2}
\end{figure}

{\it Fixed points.} -- By comparing NRG eigenvalue flows, we have
verified that for any $t \neq 0$
the model I flows to the same strong coupling FL fixed point.  The
spectrum is a combination of two FL spectra: one for odd-length and
one for even-length free electron Wilson chain \cite{wilson1975}.  Odd
channel gathers a $\pi/2$ phase shift, while even channel has zero
phase shift. Since the conductance is given by $G=G_0
\sin^2(\delta_o-\delta_e)$ with $G_0=2e^2/h$, the system is fully
conducting in its low-temperature ground state for any non-zero
$t$. Weak perturbations of the form $H'=V n_i$ are marginal, therefore
the triple quantum dot system has an extended region of high
conductance as a function of the gate voltage \cite{tripike}. The
unstable intermediate temperature fixed-point spectrum of model I is
in agreement with the conformal field theory predictions for the 2CK
model \cite{affleck2ck1992} (Fig.~\ref{fig_i2}). The same fixed point
is obtained for all $J$ in model II.

{\it Robustness of the NFL regime.} -- In model I, the channel
symmetry is broken intrinsically by the inter-impurity hopping, which
contributes a left-right cotunneling term of the form $J_{\mathrm{LR}}
\vc{S} \cdot (\vc{s}_{\mathrm{LR}} + \vc{s}_{\mathrm{RL}})$ to the
effective Hamiltonian ($\vc{s}_{\mathrm{LR}}$ is the left-right spin
operator $\vc{s}_{\mathrm{LR}} = \sum_{kk'\alpha\beta}
c^\dag_{Lk\alpha} \vc{\sigma}_{\alpha\beta} c_{Rk'\beta}$)
\cite{kuzmenko2003}. The impurities then couple to a symmetric and
antisymmetric combination of channels with exchange constants
$J_{\mathrm{S},\mathrm{A}} = J_\mathrm{avg} \pm J_\mathrm{LR}$.  The
asymmetry parameter $A=\Delta/\mathcal{J}^2$ with $\Delta =
\rho_0(J_\mathrm{S}-J_\mathrm{A})=2\rho_0 J_\mathrm{LR}$ and
$\mathcal{J}=\rho_0(J_\mathrm{S}+J_\mathrm{A})/2=\rho_0
J_\mathrm{avg}$ determines the cross-over scale $T_\Delta/T_K \approx
A^2$ \cite{pustilnik2004}. Estimating $J_S$ and $J_A$ for
$t/D=0.005$ using the Schrieffer-Wolff transformation we obtain $A^2 \sim
10^{-6}$, to be compared with $T_\Delta/T_\mathrm{scr} \sim 10^{-5}$
determined by the NRG calculation. The discrepancy appears due to
competing magnetic ordering and Kondo screening (and emerging
two-stage Kondo physics); simple scaling approach fails in this case.

We have performed a range of calculations for various perturbations
for $t/D=0.005$.
The $\ln2/2$ NFL plateau persists even for large deviations from the
p-h symmetry ($H'=V \sum_i n_i$ up to $V/U \approx 0.2$), for broken
left-right symmetry or parity ($H'=V (n_1-n_3)$ up to $V/U \approx
0.2$), and for unequal on-site repulsion parameters $U_i$. The only
``dangerous'' perturbations are those that increase the channel
asymmetry; these can be compensated in experiments by tuning on-site
energies and hybridization strengths.

\begin{figure}
\includegraphics[width=8cm,clip]{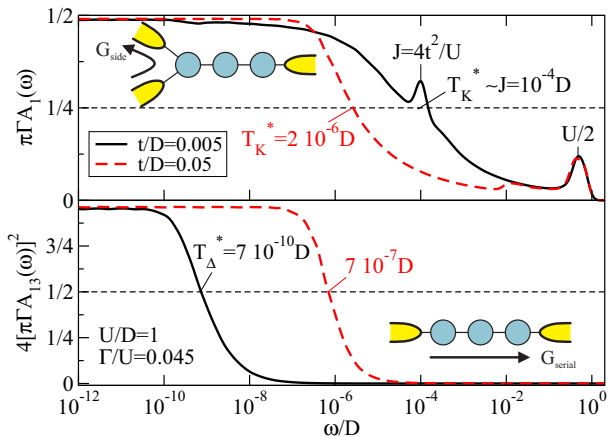}
\caption{(Color online) Dynamic properties of model I in
the AFM (dashed lines) and in the cross-over regime
(full lines). Upper panel: on-site spectral function
$A_{1}(\omega)$ of the left dot. Lower panel: out-of-diagonal
spectral function $A_{13}(\omega)$ squared. 
Temperature $T^*_\Delta$ is of order $T_\Delta$, $T_K^*$ is of
order $T_\mathrm{scr}$.
}
\label{fig2_c2}
\end{figure}

{\it Transport properties.} -- The qualitative temperature dependence
of the zero-bias conductance through the system can be inferred in a
very rough approximation from the frequency dependence of the spectral
functions. The conductance through the system is given by
$G_\mathrm{serial}/G_0 \approx 4(\pi \Gamma A_{13})^2$
\cite{caroli1971} and the conductance through a side dot in three
terminal configuration by $G_\mathrm{side}/G_0 \approx \pi \Gamma A_1$
\cite{meir1993}.  The appropriately normalized spectral densities are
shown in Fig.~\ref{fig2_c2} for the cases of cross-over regime with a
NFL region and AFM regime with no discernible NFL behavior. In the NFL
region ($t/D=0.005$ and $T_\Delta\lesssim T\lesssim T_{\mathrm{scr}}$), the
conductance $G_\mathrm{side} \sim 1/2 G_0$, while $G_\mathrm{serial}
\sim 0$. The increase of the conductance through the system at $T
\lesssim T_\Delta$ is concomitant with the cross-over from NFL to FL
fixed point, since charge transfer (or, equivalently, channel
asymmetry) destabilizes the NFL fixed point like in the two-impurity
case \cite{zarand2006}. In the AFM regime with no NFL region, both
conductances increase below the same temperature scale, {\it i.e.}
$T\lesssim T_{\mathrm{scr}}$. Measuring $G_\mathrm{side}$ and
$G_\mathrm{serial}$ could therefore serve as an experimental probe for
observation of NFL physics.

In $A_1(\omega)$, the Hubbard peak at $U/2$ corresponds to adding an
electron to the site, while the ``magnetic-excitation'' peak at $J$
appears when, after adding an electron, the electron with the opposite
spin hops from the impurity into the band. This breaks the AFM spin
chain, increasing the energy by $J$. This magnetic peak evolves into a
``molecular-orbital'' peak at the energy of the non-bonding orbital
(for $t$ in MO regime) or into the Kondo peak of the side dot (for $t$
in TSK regime). Note also the different approach to the $\omega=0$
limit in FL and NFL cases.

{\it Conclusion.} -- In a wide interval around the p-h symmetric
point, the triple quantum dot system has a FL ground state with high
conductance at $T=0$. The different regimes exhibit different
approaches to this fixed point.
The most likely candidate for observing 2CK behavior is the cross-over
regime with competing magnetic ordering and Kondo screening, $J \sim
T_K$. In this regime the NFL behavior occurs in a wide temperature
range and it is fairly robust against various perturbation that do not
additionally increase the channel asymmetry. The signature of the NFL
behavior can be detected by measuring $G_\mathrm{side}$ and
$G_\mathrm{serial} $ in a three terminal configuration. Properly
choosing parameters of the triple quantum dot system to set it into
the crossover regime represents a road map for observation of NFL
behavior.

We remark that all systems of quantum dots described as Kondo or
Anderson impurities that are coupled to two noninteracting leads
belong to the same class of quantum impurity problems with four (spin
and channel) flavors of bulk fermions \cite{affleck1995,
maldacena1997}.
The boundary conformal field theory approach allows for systematic
determination of all possible non-trivial (NFL) fixed points. It would
be interesting to investigate the {\it inverse problem}
of finding the
microscopic Hamiltonians which renormalize to these
points. Since RG transformations are not invertible, this appears as a
formidable task that would involve considerable guesswork.
Nevertheless, its resolution would provide important insight into a variety
of possible experimental realizations of mesoscopic NFL systems with {\it
purely local} interactions.

The authors acknowledge useful discussions with A. Ram\v{s}ak and
the financial support of the SRA under Grant No. P1-0044.

\bibliography{paper}

\end{document}